\documentclass{conm-p-l}

\usepackage{graphics}

\newtheorem{theorem}{Theorem}[section]
\newtheorem{lemma}[theorem]{Lemma}

\theoremstyle{definition}

\theoremstyle{remark}

\numberwithin{equation}{section}

\newcommand{\F}{{\mathcal F}}
\newcommand{\e}{\epsilon}
\newcommand{\vth}{\vartheta}

\newcommand{\D}{{\mathcal D}}
\renewcommand{\k}{\kappa}

\newcommand{\hS}{\hat{S}}
\newcommand{\tS}{\tilde{S}}

\newcommand{\Dl}{\Delta}
\renewcommand{\th}{\theta}

\newcommand{\ra}{\rightarrow}

\newcommand{\sg}{\sigma}

\newcommand{\pa}{\partial}

\newcommand{\La}{\Lambda}

\newcommand{\la}{\lambda}

\newcommand{\nid}{\noindent}

\newcommand{\om}{\omega}
\newcommand{\Om}{\Omega}

\renewcommand{\O}{{\mathcal O}}
\newcommand{\non}{\nonumber}

\begin{document}

\title{Homoclinic Tubes and Chaos in Perturbed Sine-Gordon Equation}

\author{Y. Charles  Li}
\address{Department of Mathematics, University of Missouri, 
Columbia, MO 65211}
\curraddr{}
\email{cli@math.missouri.edu}
\thanks{}

%\author{Author Two}
%\address{Mathematical Research Section, School of Mathematical Sciences,
%Australian National University, Canberra ACT 2601, Australia}

%    General info
\subjclass{35, 37, 34, 78}
\date{}



\begin{abstract}
In \cite{Li03b}, Bernoulli shift dynamics of submanifolds was 
established in a neighborhood of a homoclinic tube. In this article, 
we will present a concrete example: sine-Gordon equation under a 
quasi-periodic perturbation.
\end{abstract}

\maketitle

%\section*{}
%This is an example of an unnumbered first-level heading.

%\specialsection*{This is a Special Section Head}
%This is an example of a special section head%
%%%%%%%%%%%%%%%%%%%%%%%%%%%%%%%%%%%%%%%%%%%%%%%%%%%%%%%%%%%%%%%%%%%%%%%%
%\footnote{Here is an example of a footnote. Notice that this footnote
%text is running on so that it can stand as an example of how a footnote
%with separate paragraphs should be written.
%\par
%And here is the beginning of the second paragraph.}%
%%%%%%%%%%%%%%%%%%%%%%%%%%%%%%%%%%%%%%%%%%%%%%%%%%%%%%%%%%%%%%%%%%%%%%%%

%\section{This is a numbered first-level section head}
%This is an example of a numbered first-level heading.

%\subsection{This is a numbered second-level section head}
%This is an example of a numbered second-level heading.

%\subsection*{This is an unnumbered second-level section head}
%This is an example of an unnumbered second-level heading.

%\subsubsection{This is a numbered third-level section head}
%This is an example of a numbered third-level heading.

%\subsubsection*{This is an unnumbered third-level section head}
%This is an example of an unnumbered third-level heading.

\section{Introduction}

Propagations of nonlinear waves through homogeneous media are often 
modeled by well-known nonlinear wave equations, for example, sine-Gordon 
equation. Studies were also drawn to variable media \cite{ZKV92} 
\cite{GSW01} \cite{CM95}. Variations of the media can speed up, slow 
down (even stop), or break the wave propagations. Studies have been 
focused upon such variations of the media, which are localized defects. 
In the current article, we will study quasi-periodic media. The equation 
to be studied can be called a quasi-periodically defective sine-Gordon 
equation. This equation represents a concrete example realizing the 
theorem proved in \cite{Li03b} \cite{Li03a}. Consequently, existence of 
a homoclinic tube asymptotic to a torus can be proved, and Bernoulli 
shift dynamics of tori can be established.

The article is organized as follows: In section 2, we present the 
formulation of the problem. Section 3 is on an intergable theory. 
Section 4 is on the existence of a homoclinic tube and chaos.

\section{Formulations of the Problem}

Consider the sine-Gordon equation under a quasi-periodic perturbation,
\begin{equation}
u_{tt}=c^2 u_{xx} + \sin u +\e [ \D u + f(t) (\sin u - u)]\ ,
\label{PSG}
\end{equation}
which is subject to periodic boundary condition and odd constraint
\begin{equation}
u(t, x+2\pi ) = u(t, x)\ , \quad u(t, -x) = - u(t, x)\ ,
\label{obc}
\end{equation}
where $u$ is a real-valued function of two real variables $t \geq 0$ 
and $x$, c is a parameter, $\frac{1}{2} < c < 1$, $\e$ is a small 
parameter, $\e \geq 0$, $\D$ is a dissipative operator
\[
\D u = -a u_t + b u_{txx}\ , \quad a \geq 0\ , \quad b \geq 0\ ,
\]
and $f(t)$ is quasi-periodic with a basis of frequencies $\om_1, \cdots, 
\om_N$, 
\[
f(t)=\sum_{n=1}^N a_n \cos [\th_n(t)]\ , \quad \th_n(t)=\om_n t + \th_n^0\ ,
\]
where $a_n$ and $\th_n^0$ are parameters. The above system is invariant 
under the transform $u \ra -u$. On the other hand, the odd constraint 
prohibits the transform $u \ra u + 2\pi$. Equation (\ref{PSG}) is 
equivalent to the following system
\begin{equation}
\left ( \begin{array}{c} u \cr v \cr \end{array} \right )_t 
= L \left ( \begin{array}{c} u \cr v \cr \end{array} \right ) +
\left ( \begin{array}{c} 0 \cr \sin u +\e f(t) (\sin u - u) \cr 
\end{array} \right )\ , 
\label{EPSG}
\end{equation}
where
\[
L \left ( \begin{array}{c} u \cr v \cr \end{array} \right ) = 
\left ( \begin{array}{c} v \cr c^2 u_{xx} -\e a v +\e b v_{xx} \cr 
\end{array} \right )\ .
\]
When $\e = 0$, $L$ generates a $C_0$ semi-group on $H^1 \times L^2$ 
(the Sobolev spaces $H^1$ and $H^0=L^2$ on $[0,2\pi]$), and the domain 
of $L$ is $H^2 \times H^1$. When $\e \neq 0$, $L$ still generates a 
$C_0$ semi-group on $H^1 \times L^2$, but if $b \neq 0$, then the domain 
of $L$ is $H^2 \times H^2$. Since the nonlinear term in (\ref{EPSG}) is 
uniformly Lipschitz, (\ref{EPSG}) is globally well-posed in 
$C([0, \infty ), H^1 \times L^2)$. That is, for any $(u_0,v_0) \in H^1 
\times L^2$, there exists a unique mild solution $(u(t),v(t)) \in 
C([0, \infty ), H^1 \times L^2)$ such that $(u(0),v(0)) = (u_0,v_0)$. 
One can introduce the evolution operator $F^t$ as $(u(t),v(t))=
F^t(u_0,v_0)$. If $b = 0$, $F^t$ is defined for all $t \in \mathbb{R}$, 
and for any fixed $t$, $F^t$ is a $C^\infty$ diffeomorphism. For a 
classical reference, see \cite{Paz83}.

\section{Integrable Theory}

When $\e =0$, equation (\ref{PSG}) reduces to the well-known sine-Gordon 
equation 
\begin{equation}
u_{tt}=c^2 u_{xx} + \sin u\ , 
\label{SG}
\end{equation}
which is integrable through the Lax pair
\begin{eqnarray}
\psi_x &=& B\psi\ , \label{Lax1} \\
\psi_t &=& A\psi\ , \label{Lax2}
\end{eqnarray}
where
\[
B= \frac{1}{c} \left ( \begin{array}{lr}\frac{i}{4}(cu_x +u_t) & 
\frac{1}{16\la}e^{iu}+\la \cr \cr -\frac{1}{16\la}e^{-iu}-\la & -
\frac{i}{4}(cu_x +u_t) \cr \end{array}\right )\ , 
\]
\[
A= \left ( \begin{array}{lr}\frac{i}{4}(cu_x +u_t) & -
\frac{1}{16\la}e^{iu}+\la \cr \cr \frac{1}{16\la}e^{-iu}-\la & -
\frac{i}{4}(cu_x +u_t) \cr \end{array}\right )\ . 
\]
The Lax pair (\ref{Lax1})-(\ref{Lax2}) possesses a symmetry.
\begin{lemma}
If $\psi = \left ( \begin{array}{c}\psi_1 \cr \psi_2 \cr \end{array}
\right )$ solves the Lax pair (\ref{Lax1})-(\ref{Lax2}) at $(\la, u)$, 
then $\left ( \begin{array}{c}\overline{\psi_2} \cr \cr \overline{\psi_1} 
\cr \end{array}\right )$ solves the Lax pair (\ref{Lax1})-(\ref{Lax2}) 
at $(-\bar{\la}, u)$.
\label{symm}
\end{lemma}
\nid
There is a Darboux transformation for the Lax pair (\ref{Lax1})-(\ref{Lax2}).
\begin{theorem}[Darboux Transformation I] Let
\begin{eqnarray}
U &=& u +2i \ln \left [ \frac{i\phi_2}{\phi_1} \right ]\ , \non \\
\Psi &=& \left ( \begin{array}{lr} -\nu \phi_2/\phi_1 & \la \cr 
-\la & \nu \phi_1/\phi_2 \cr \end{array}\right ) \psi \ , \non 
\end{eqnarray}
where $\phi = \psi |_{\la = \nu}$ for some $\nu$, then $\Psi$ solves the 
Lax pair (\ref{Lax1})-(\ref{Lax2}) at $(\la, U)$.
\label{DT1}
\end{theorem}
Often in order to guarantee the reality condition (i.e. $U$ needs to be 
real-valued), one needs to iterate the Darboux transformation by virtue 
of Lemma \ref{symm}. The result corresponds to the counterpart of the 
Darboux transformation for the cubic nonlinear Schr\"odinger equation 
\cite{LM94}.
\begin{theorem}[Darboux Transformation II] Let
\begin{eqnarray}
U &=& u +2i \ln \left [ \frac{\nu |\phi_1|^2 +\bar{\nu}|\phi_2|^2}
{\bar{\nu} |\phi_1|^2 +\nu|\phi_2|^2} \right ]\ , \non \\
\Psi &=& G \psi \ , \non
\end{eqnarray}
where 
\begin{eqnarray}
G &=& \left ( \begin{array}{lr}G_1 & G_2 \cr G_3 & G_4 \cr \end{array}
\right ) \ , \non \\
G_1 &=& |\nu|^2 \frac{\nu |\phi_1|^2 +\bar{\nu}|\phi_2|^2}
{\bar{\nu} |\phi_1|^2 +\nu|\phi_2|^2} - \la^2 \ , \non \\
G_2 &=& \frac{\la (\nu^2-\bar{\nu}^2) \phi_1 \overline{\phi_2}}
{\bar{\nu} |\phi_1|^2 +\nu|\phi_2|^2} \ , \non \\
G_3 &=& \frac{\la (\nu^2-\bar{\nu}^2) \overline{\phi_1}\phi_2}
{\nu |\phi_1|^2 +\bar{\nu}|\phi_2|^2} \ , \non \\
G_4 &=& |\nu|^2 \frac{\bar{\nu} |\phi_1|^2 +\nu|\phi_2|^2}{\nu |\phi_1|^2 
+\bar{\nu}|\phi_2|^2} - \la^2 \ , \non 
\end{eqnarray}
and $\phi = \psi |_{\la = \nu}$ for some $\nu$, then $\Psi$ solves the 
Lax pair (\ref{Lax1})-(\ref{Lax2}) at $(\la, U)$.
\label{DT2}
\end{theorem}
Proof: Let $\phi$ be an eigenfunction solving the Lax pair 
(\ref{Lax1})-(\ref{Lax2}) at $(\la, u)$. With ($\phi, \nu, u$), the 
Darboux transformation given in Theorem \ref{DT1} leads to 
\begin{eqnarray}
\tilde{U} &=& u +2i \ln \left [ \frac{i\phi_2}{\phi_1} \right ]\ , 
\label{pdt1} \\
\tilde{\Psi} &=& \left ( \begin{array}{lr} -\nu \phi_2/\phi_1 & \la \cr 
-\la & \nu \phi_1/\phi_2 \cr \end{array}\right ) \psi \ . \label{pdt2}
\end{eqnarray}
By Lemma \ref{symm}, $\hat{\phi} =\left ( \begin{array}{c}
\overline{\phi_2} \cr \cr \overline{\phi_1} \cr \end{array}\right )$ 
solves the Lax pair (\ref{Lax1})-(\ref{Lax2}) at $(-\bar{\nu}, u)$. Hence,
\[
\hat{\Phi} = \left ( \begin{array}{lr} -\nu \phi_2/\phi_1 & -
\bar{\nu} \cr 
\bar{\nu} & \nu \phi_1/\phi_2 \cr \end{array}\right )\hat{\phi}
\]
solves the Lax pair (\ref{Lax1})-(\ref{Lax2}) at $(-\bar{\nu}, \tilde{U})$. 
With ($\hat{\Phi}, -\bar{\nu}, \tilde{U}$), the Darboux transformation given 
in Theorem \ref{DT1} leads to the expressions given in the current 
theorem. Q.E.D.

Focusing upon the spatial part (\ref{Lax1}) of the Lax pair, one can 
develop a complete Floquet theory. Let $M(x)$ be the fundamental matrix 
of (\ref{Lax1}), $M(0)=I$ ($2\times 2$ identity matrix), then the Floquet 
discriminant is given as 
\[
\Dl = \ \mbox{trace} \ M(2\pi)\ .
\]
The Floquet spectrum is given by
\[
\sg = \{ \la \in \mathbb{C}\ | \ -2 \leq \Dl(\la) \leq 2 \} \ .
\]
Periodic and anti-periodic points $\la^{\pm}$ (which correspond to 
periodic and anti-periodic eigenfunctions respectively) are defined by
\[
\Dl(\la^{\pm}) = \pm 2 \ .
\]
A critical point $\la^{(c)}$ is defined by
\[
\frac{d\Dl}{d\la}(\la^{(c)}) = 0 \ .
\]
A multiple point $\la^{(m)}$ is a periodic or anti-periodic point which 
is also a critical point. The algebraic multiplicity of $\la^{(m)}$ is 
defined as the order of the zero of $\Dl(\la)\pm 2$ at $\la^{(m)}$. When 
the order is $2$, we call the multiple point a double point, and denote 
it by $\la^{(d)}$. The order can exceed $2$. The geometric multiplicity 
of $\la^{(m)}$ is defined as the dimension of the periodic or 
anti-periodic eigenspace at $\la^{(m)}$, and is either $1$ or $2$.   

Counting lemmas for $\la^{\pm}$ and $\la^{(c)}$ can be established 
similarly as in \cite{LM94}. As a result, there exist sequences 
$\{ \la^{\pm}_j \}$ and $\{ \la^{(c)}_j \}$. An important sequence of 
invariants $F_j$ of the sine-Gordon equation can be defined by
\[
F_j(u,u_t)= \Dl(\la^{(c)}_j(u,u_t),u,u_t)\ .
\]
If $\{ \la^{(c)}_j \}$ is a simple critical point of $\Dl$, then
\[
\frac{\pa F_j}{\pa w} = \frac{\pa \Dl}{\pa w}\bigg |_{\la =\la^{(c)}_j}\ , 
\quad w=u, u_t\ .
\]
As a function of three variables, $\Dl = \Dl(\la, u, u_t)$ has the 
partial derivatives given by Bloch functions $\psi^\pm$ (i.e. $\psi^\pm (x) 
= e^{\pm \La x}\tilde{\psi}^\pm (x)$, where $\tilde{\psi}^\pm$ are 
periodic in $x$ of period $2\pi$, and $\La$ is a complex constant):
\begin{eqnarray}
\frac{\pa \Dl}{\pa u} &=& \frac{-i}{16\la c} \frac{\sqrt{\Dl^2-4}}
{W(\psi^+,\psi^-)} \bigg [ 4\la c \pa_x (\psi^+_1\psi^-_2 +\psi^+_2
\psi^-_1) 
+ e^{-iu} \psi^+_1\psi^-_1 -e^{iu} \psi^+_2\psi^-_2 \bigg ]\ , \non \\
\frac{\pa \Dl}{\pa u_t} &=& \frac{i}{4 c} \frac{\sqrt{\Dl^2-4}}
{W(\psi^+,\psi^-)} \bigg [ \psi^+_1\psi^-_2 +\psi^+_2\psi^-_1 \bigg ] \ , 
\non \\
\frac{\pa \Dl}{\pa \la } &=& \frac{-1}{c} \frac{\sqrt{\Dl^2-4}}
{W(\psi^+,\psi^-)} \int_0^{2\pi}\bigg [ \bigg ( \frac{1}{16\la^2}e^{iu} -1 
\bigg )\psi^+_2\psi^-_2 + \bigg ( \frac{1}{16\la^2}e^{-iu} -1 \bigg )
\psi^+_1\psi^-_1 \bigg ] dx \ , \non
\end{eqnarray}
where ${W(\psi^+,\psi^-)}= \psi^+_1\psi^-_2 -\psi^+_2\psi^-_1$ is the 
Wronskian. Of course, the sine-Gordon equation can be written in the 
Hamiltonian form
\[
u_t = \frac{\pa H}{\pa v}\ , \quad v_t = -\frac{\pa H}{\pa u}\ ,
\]
where the Hamiltonian is given by 
\[
H = \int_0^{2\pi}\bigg [ \frac{1}{2} (v^2+c^2u_x^2)+\cos u \bigg ] dx\ .
\]
It turns out that $F_j$'s provide the perfect Melnikov vectors rather 
than the Hamiltonian or other invariants \cite{LM94}.

\begin{figure}
\includegraphics{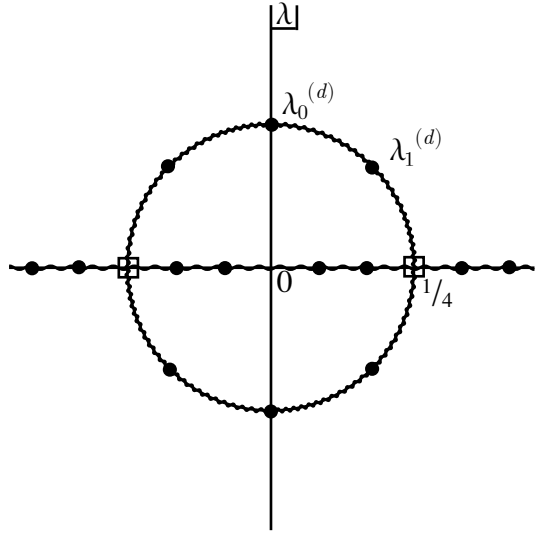}
\caption{Floquet spectrum of the Lax pair at $u=0$, $\bullet$ double 
point, $\square$ critical point.}
\label{fsp}
\end{figure}

$u=0$ is a fixed point of the sine-Gordon equation. Linearization of the 
sine-Gordon equation at $u=0$ leads to 
\[
u_{tt}=c^2u_{xx} +u \ .
\]
Let $u = \sum_{k=1}^\infty u_k^0 e^{\Om_k t} \sin kx$, $u_k^0$ and $\Om_k$ 
are constants, then 
\[
\Om_k = \pm \sqrt{1-c^2k^2}\ , \quad k=1,2\cdots .
\]
Since $1/2 < c <1$, only $k=1$ is an unstable mode, the rest modes are 
neutrally stable. The corresponding nonlinear unstable foliation can be 
represented through the Darboux transformation given in Theorem \ref{DT2}. 
When $u=0$, the Bloch functions of the Lax pair (\ref{Lax1})-(\ref{Lax2}) 
are 
\[
\psi^{\pm} = e^{\pm i (\k x +\om t)} \left ( \begin{array}{c} 1 \cr 
\pm i \cr \end{array}\right )\ , 
\]
where $\k = \frac{1}{c}(\la + \frac{1}{16\la })$ and $\om = \la - 
\frac{1}{16\la }$. Thus,
\[
\la = \frac{1}{2} \bigg [ \k c \pm \sqrt{(\k c)^2 -\frac{1}{4}}\bigg ]\ .
\]
The Floquet discriminant is given by
\[
\Dl = 2 \cos (2\pi \k)\ .
\]
The spectral data are depicted in Figure \ref{fsp}.
Noticing that the Darboux transformation in Theorem \ref{DT2} depends upon 
quadratic products of eigenfunctions, one realizes that $\phi$ should be 
chosen at $\k = \frac{1}{2}$ and $\nu = \frac{1}{4} [c+i\sqrt{1-c^2}]$. 
$\nu$ is a complex double point, $\nu = \la_1^{(d)}$ in Figure \ref{fsp}. 
(It turns out that for other soliton equations , e.g. Davey-Stewartson II 
equation \cite{Li99}, 
$\nu$ may not be a double point.) The wise choice for $\phi$ is 
\[
\phi = \sqrt{\frac{c_+}{c_-}} \phi^+ + \sqrt{\frac{c_-}{c_+}} \phi^-\ ,
\]
where
\[
\phi^{\pm} = e^{\pm i\frac{1}{2} x \mp \frac{\sg }{2}t} \left ( 
\begin{array}{c} 1 \cr 
\pm i \cr \end{array}\right )\ , \quad \sg = \sqrt{1-c^2}\ , 
\]
and $c_\pm$ are arbitrary complex constants. Let 
\[
\frac{c_+}{c_-}=e^{\rho +i\th }\ , \quad \tau = \sg t - \rho \ , \quad 
\xi =x +\th \ ,
\]
then 
\[
\phi = 2 \left ( \begin{array}{c} \cosh \frac{\tau }{2} \cos \frac{\xi }{2}
-i \sinh \frac{\tau }{2} \sin \frac{\xi }{2} \cr 
-\cosh \frac{\tau }{2} \sin \frac{\xi }{2}
-i \sinh \frac{\tau }{2} \cos \frac{\xi }{2} \cr
\end{array}\right )\ .
\]
The Darboux transformation in Theorem \ref{DT2} leads to 
\begin{equation}
U = \pm 4 \vth\ , \quad \vth = \arctan \bigg [ \frac{\sg }{c} \ \mbox{sech} 
\tau \sin x \bigg ] \ , \quad  \vth \in (-\frac{\pi }{2}, \frac{\pi }{2})\ , 
\label{horbit}
\end{equation}
corresponding to $\th = \pm \frac{\pi }{2}$ (which in turn corresponds 
to the $U \ra -U$ symmetry). Notice that
\[
\mbox{det}\ G = (\la^2-\nu^2)(\la^2-\bar{\nu}^2)\ ,
\]
L'Hospital's rule implies that 
\[
\lim_{\la \ra \nu } \frac{\sqrt{\Dl^2-4}}{W(\Psi^+,\Psi^-)} 
=\frac{\sqrt{\Dl(\nu )\Dl''(\nu )}}{2\nu (\nu^2-\bar{\nu}^2)
W(\phi^+,\phi^-)}\ .
\]
Moreover,
\begin{eqnarray}
\Phi^\pm &=& G \phi^\pm = \pm e^{\mp (\frac{\rho}{2} +i \frac{\th}{2})}
\nu (\nu^2-\bar{\nu}^2)W(\phi^+,\phi^-) \non \\
& & \times \left (\begin{array}{c} \frac{-\overline{\phi_2}}{\bar{\nu}
|\phi_1|^2+\nu |\phi_2|^2} \cr \cr \frac{\overline{\phi_1}}{\nu |\phi_1|^2+
\bar{\nu} |\phi_2|^2} \cr \end{array} \right )\ . \non
\end{eqnarray}
Finally,
\begin{eqnarray}
\frac{\pa F_1}{\pa u_t}\bigg |_{u=U} &=& \frac{\pi \sg^2}{\sqrt{2}c}\ 
\mbox{sech} \tau [c^2+\sg^2\ \mbox{sech}^2 \tau \sin^2 x ]^{-1} \non \\
& & \times [\pm \tanh \tau \sin x \mp i \cos x ] \ , \label{melv}
\end{eqnarray}
corresponding to $\th =\pm \frac{\pi}{2}$. The real part of (\ref{melv}) 
is the Melnikov vector. 

\section{Existence of a Homoclinic Tube and Chaos}

The defective sine-Gordon equation (\ref{PSG}) can be related to an 
autonomous system by introducing extra phase variables 
$\th = (\th_1, \cdots, \th_N)$,
\begin{eqnarray}
& & u_{tt}=c^2 u_{xx} + \sin u +\e [ \D u + f(\th_1, \cdots, \th_N) 
(\sin u - u)]\ , \label{APSG1} \\
& & \frac{d\th_n}{dt} = \om_n\ , \quad (n=1,\cdots,N)\ . \label{APSG2}
\end{eqnarray}
For any $\th^0$, solving (\ref{APSG2}), equation (\ref{APSG1}) becomes 
(\ref{PSG}).

$u=0$ corresponds to a $N$-torus denoted by $\tS$. Linearization at 
$u=0$ leads to 
\begin{eqnarray}
& & u_{tt}=c^2 u_{xx} + u +\e \D u \ , \non \\
& & \frac{d\th_n}{dt} = \om_n\ , \quad (n=1,\cdots,N)\ . \non
\end{eqnarray}
Thus $u=0$ corresponds to a normally hyperbolic $N$-torus with one 
unstable mode (since $1/2 < c<1$), when $\e > 0$. Proofs of the following 
invariant manifold theorem have become standard after the works 
\cite{LMSW96} \cite{Li03c}.
\begin{theorem}
The $N$-torus $\tS$ has an ($N+1$)-dimensional $C^m$ ($m \geq 3$) 
center-unstable manifold $W^{cu}$  and a $1$-codimensional $C^m$ 
center-stable manifold $W^{cs}$ in the phase space $(u,u_t,\th) \in H^1 
\times L^2 \times \mathbb{T}^N$. $W^{cu} \cap W^{cs} = \tS$. $W^{cu}$ 
is $C^1$ in $\e$ for $\e \in [0,\e_0)$ and some $\e_0 >0$. When $b \neq 0$, 
for $(u,u_t) \in H^2\times H^2$, $W^{cs}$ is $C^1$ in $\e$ for $\e \in 
[0,\e_0)$. When $b = 0$, $W^{cs}$ is always $C^1$ in $\e$ for $\e \in 
[0,\e_0)$. Inside $W^{cu}$ and $W^{cs}$ respectively, there are a $C^m$ 
invariant family of $1$-dimensional $C^m$ unstable fibers $\{ \F^u(\th):\ 
\th \in \tS \}$ and a $C^m$ invariant family of $C^m$ stable fibers 
$\{ \F^s(\th):\ \th \in \tS \}$, such that
\[
W^{cu} = \bigcup_{\th \in \tS} \F^u(\th)\ , \quad W^{cs} = 
\bigcup_{\th \in \tS} \F^s(\th)\ .
\]
There are positive constants $\k_u = \frac{1}{2} \sqrt{1-c^2}$, 
$\k_s= \frac{1}{4}\e(a+b)$, and $C$ such that 
\begin{eqnarray}
\| F^t(q^-)-F^t(\th)\| &\leq& Ce^{\k_ut}\| q^- -\th \| \ , \quad \forall 
t \in (-\infty, 0]\ , \label{decay1} \\
& & \forall \th \in \tS\ , \quad \forall q^- \in 
\F^u(\th)\ , \non \\
\| F^t(q^+)-F^t(\th)\| &\leq& Ce^{-\k_st}\| q^+ -\th \| \ , \quad \forall 
t \in [0, +\infty)\ , \label{decay2} \\ 
& & \forall \th \in \tS\ , \quad \forall q^+ \in 
\F^s(\th)\ , \non \\ 
\| F^t(\th^+)-F^t(\th^-)\| &\leq& C\| \th^+ -\th^- \| \ , \quad \forall 
t \in (-\infty, +\infty)\ , \non \\ 
& & \forall \th^+, \th^- \in \tS\ , \non
\end{eqnarray}
where $F^t$ is the evolution operator of (\ref{APSG1})-(\ref{APSG2}).
\label{invthm}
\end{theorem}
In terms of the original setting (\ref{PSG}), $\F^u(\th)$ and $\F^s(\th)$
are the unstable and stable manifolds of the fixed point $u=0$, which are 
$C^m$ smooth in $\th^0$. As shown in \cite{LMSW96} \cite{Li03c}, to the 
leading order, the signed distance between $\F^u(\th)$ and $\F^s(\th)$ 
(which is a certain coordinate difference) is given by the Melnikov integral
\[
M = \int_{-\infty}^{+\infty} \int_0^{2\pi} \bigg \{ \frac{\pa F_1}{\pa u_t} 
\bigg [ \D u + f(t) (\sin u - u)\bigg ] \bigg \}_{u = U} dx dt \ ,
\]
where $U$ is given in (\ref{horbit}) and $\frac{\pa F_1}{\pa u_t}
|_{u = U}$ is given in (\ref{melv}). The signed distance between 
$\F^u(\th)$ and $\F^s(\th)$ is $C^m$ ($m \geq 3$) in $\th^0$. The zero of 
the Melnikov integral and implicit function theorem imply the following 
theorem, for detailed arguments, see \cite{LMSW96} \cite{Li03c}.
\begin{theorem}
If $b \neq 0$, there is a region for $(a,b)$ in $\mathbb{R}^+ \times 
\mathbb{R}^+$, or if $b =0$, there is a region for $a$ in $\mathbb{R}^+$, 
such that $W^{cu}$ and $W^{cs}$ intersect into a $N$-dimensional $C^m$ 
($m\geq 3$) homoclinic tube $\hS$ asymptotic to the $N$-torus $\tS$. 
\end{theorem}
Additional remarks for the proof of the theorem are that the size of 
$\F^s(\th)$ is of order $\O(\sqrt{\e})$ since the nonlinear term in 
(\ref{PSG}) is cubic. Therefore, the so-called second measurement in 
\cite{LMSW96} \cite{Li03c} is not needed. 

The rest of this article only deals with the case $b=0$. 
The Poincar\'e period map $F$ determined by setting $\th_1 = 2n\pi$ has 
the homoclinic tube $\xi = (\cdots S_{-1} S_0 S_1 \cdots )$ which is 
asymptotic to the ($N-1$)-torus $S$ obtained from $\tS$ by setting 
$\th_1 =0$. $S_0$ is a $C^m$ ($N-1$)-torus as a result of the smoothness 
of the signed distance with respect to $\th^0$, and $S_j=F^j S_0$, 
$\forall j \in \mathbb{Z}$. The rest of Assumption (A1) in \cite{Li03b} 
can be verified by noticing that the decay rates in 
(\ref{decay1})-(\ref{decay2}) are uniform with respect to $\th$, and the 
fact that $\F^u(\th)$ and $\F^s(\th)$ are the unstable and stable manifolds 
of the fixed point $u=0$ of (\ref{PSG}). Since $S$ is a finite-dimensional 
torus, $\xi \cup S$ is compact, thus Assumption (A2) in \cite{Li03b} is 
also satisfied. Therefore, we have the following theorem.
\begin{theorem}[Chaos Theorem]
When $b=0$, 
there is a Cantor set $\Xi$ of tori which is invariant under the iterated 
Poincar\'e map $F^{2K+1}$ for some $K$. The action of $F^{2K+1}$ on $\Xi$ 
is topologically conjugate to the action of the Bernoulli shift on two 
symbols $0$ and $1$. 
\end{theorem}

Acknowledgement: I would like to thank Brenda Frazier for artist work.

\end{document}